\documentclass[final]{raa}            % referee version: for submission
\usepackage{graphicx,times}
\usepackage{natbib}

\providecommand{\bm}[1]{\mbox{\boldmath$#1$\unboldmath}}

\begin{document}

 \title{Submillimeter/millimeter observations of
 the high-mass star forming region IRAS
   22506+5944
 $^*$ \footnotetext{\small $*$ Supported by the National Natural
Science Foundation of China.} }

   \volnopage{ {\bf 2010} Vol.\ {\bf 10} No. {\bf 2}, 151 -- 158}
   \setcounter{page}{151}

   \author{Jin-Long Xu
      %\inst{1,2}
     \and Jun-Jie  Wang
      %\inst{1,2}
      }

    \institute{National Astronomical Observatories, Chinese Academy of Sciences,
             Beijing 100012, China; {\it  xujl@bao.ac.cn} \\
             %\and
              NAOC-TU Joint Center for Astrophysics, Lhasa 850000, China \\
\vs \no
   {\small Received 2009 May 5; accepted 2009 October 24 }
}

 \abstract{ The mapping observations of CO $J=2-1$, CO
$J=3-2$, $^{13}$CO $J=2-1$ and $^{13}$CO $J=3-2$ lines {in the
direction of} IRAS~22506+5944 have been made.  The results show
that the cores in {the} $J=2-1$ transition lines have a similar
morphology to {those} in {the} $J=3-2$ transition lines. Bipolar
molecular outflows are verified. The prior IRAS~22506+5944
observations indicated that two IRAS sources and three H$_{2}$O
masers were located close to the peak position of the core. One of
{the} IRAS sources may be the driving source of the outflows. In
addition, the H$_{2}$O masers may occur in relatively warm
environments. The parameters of the dense core and outflow,
obtained by the LTE method, indicate that IRAS~22506+5944 is a
high-mass star formation region.
   }

   \authorrunning{J. L. Xu \&  J. J. Wang  }            %author_head in even pages
   \titlerunning{Submillimeter/Millimeter Observations  of  IRAS 22506+5944 }  % title_head in odd pages

   \maketitle
%
%
%________________________________________________ sections below
%
\section{Introduction}           %% first-level sections will be auto-capitalized
\label{sect:intro}

Low-mass stars in molecular clouds are formed through the
processes of collapse, accretion and outflow (Shu et al. 1987).
However, the processes for high-mass stars remain unclear because
of {the} short evolution timescales and large distances of
high-mass stars. Observational evidence suggests that young
high-mass stars usually form in a cluster environment (Lada et al.
1993). Bipolar molecular outflows are commonly found around young
high-mass stars (Shepherd \& Churchwell 1996; Zhang et al. 2001).
These outflows are generally much more massive and energetic than
those from low-mass stars. Recent surveys with single-dish
telescopes show that massive outflows are commonly associated with
{UC HII} regions (Shepherd \& Churchwell 1996; Zhang et al. 2001)
and H$_{2}$O masers (Elitzur et al. 1989).  The H$_{2}$O masers
may originate in hot cores, and be excited by the shocks
associated with outflows. However, how the outflows are driven
{is} poorly understood.

IRAS 22506+5944 has been proposed {as a} precursor of {a} UC HII
region (Molinari et al. 1998). Two of the three H$_{2}$O masers in
IRAS~22506+5944 are found close to the outflow peak position
(Jenness et al. 1995). Wang (1997) imaged this source in the
near-infrared $J$, $H$ and $K$ bands and found a star cluster
within the core. It has magnitudes of 17.7, 14.5 and 11.7,
respectively. Moreover, Zhang et al. (2001) detected bipolar
outflows from {an }observation of CO $J=2-1$ in this region. The
bipolar molecular outflows are confirmed from observation of {the}
CO $J=2-1$ line (Wu et al. 2005). In addition, Su et al. (2004)
reported interferometric observations in CO $J=1-0$, $^{13}$CO
$J=1-0$, C$^{18}$O $J=1-0$ and{ the} 3\,mm continuum, as well as
single-dish observations in CO $J=1-0$ and $^{13}$CO $J=1-0$.
{The} IRAS~22506+5944 source was identified as the outflow driving
source.

In this paper, we report the first mapping observations of
IRAS~22506+5944 in the CO $J=3-2$, $^{13}$CO $J=2-1$ and $^{13}$CO
$J=3-2$ lines, as well as the mapping observation in the CO
$J=2-1$ line. Due to the observed molecular lines at higher
frequencies, we can attain higher angular resolution, which is
critical {for} reduc{ing} the beam dilution and to identify {the}
relatively compact core and outflows. Also, higher $J$ transitions
are relatively more sensitive to hot gases. Such hot gas{es} are
more likely to be physically associated with high-mass young
stellar object{s} (Wu et al. 2005).

\section{Observations}

 \label{sect:Obs} The mapping observations of IRAS~22506+5944
 (R.A{.}(B1950) $=22^{\rm h}50^{\rm m}38.7^{\rm s}$,
Dec(B1950) $=59\dg44'58^{\prime\prime}$) {were} made in the CO
$J=2-1$, CO $J=3-2$, $^{13}$CO $J=2-1$ and $^{13}$CO $J=3-2$ lines
using the KOSMA 3\,m telescope at Gornergrat, Switzerland in April
2004. The half-power beam widths of the telescope at observing
frequencies of 230.538\,GHz, 345.789\,GHz, 220.399\,GHz and
330.588\,GHz are $130^{\prime\prime}$, $80^{\prime\prime}$,
$130^{\prime\prime}$ and $80^{\prime\prime}$, respectively. The
pointing and tracking accuracy is better than $10^{\prime\prime}$.
{The} DSB receiver noise temperature was about 120\,K. The medium
and variable resolution acousto{-}optical spectrometers have 1501
and 1601 channels, with total bandwidth{s} of 248\,MHz  and
544\,MHz, and equivalent velocity resolution{s} of 0.21\,${\rm km\
s^{-1}}${ and} 0.29\,${\rm km\ s^{-1}}$, respectively. The beam
efficiency $B_{\rm eff}$ is 0.68 at 230\,GHz and 220\,GHz. The
beam efficiency $B_{\rm eff}$ is 0.72 at 330\,GHz and 345\,GHz.
The fo{r}ward efficiency $F_{\rm eff}$ is 0.93. The mapping {was
done} using on-the-fly mode with {a} $1^{\prime}\times 1^{\prime}$
grid. The data were reduced using the CLASS (Continuum and Line
Analysis Single-Disk Software) and GREG (Grenoble Graphic)
software. The $80^{\prime\prime}$ resolution of {the} CO $J=3-2$
and $^{13}$CO $J=3-2$ data {was} convolved to 130$^{\prime\prime}$
with an effective beam size of
$\sqrt{130^{2}-80^{2}}=102^{\prime\prime}$. The correction for the
line intensities to {the }main beam temperature scale was made
using the formula $T_{\rm mb}= (F_{\rm eff}/B_{\rm eff}\times
T^\ast_{\rm A})$.

\section{Results and analysis}
\label{sect:data}
\subsection{ The Molecular Line Spectra }
Figure~1 gives the spectra of the different transition lines of CO
isotopes at the IRAS~22506+5944 position. Each spectrum shows
broad line wings. The blue and red wings are asymmetrical and the
spectral profiles are not Gaussian shape{d}. The observed
parameters of the IRAS~22506+5944 source are summarized in
Table~1. %11111111111111
 In Table~1, the larger full widths (FW)
appear to indicate high-velocity gas motion in this region. The
ranges of full widths are determined from the PV diagram.

\begin{figure}[h!!]

\vs
   \centering
   \includegraphics[width=55mm]{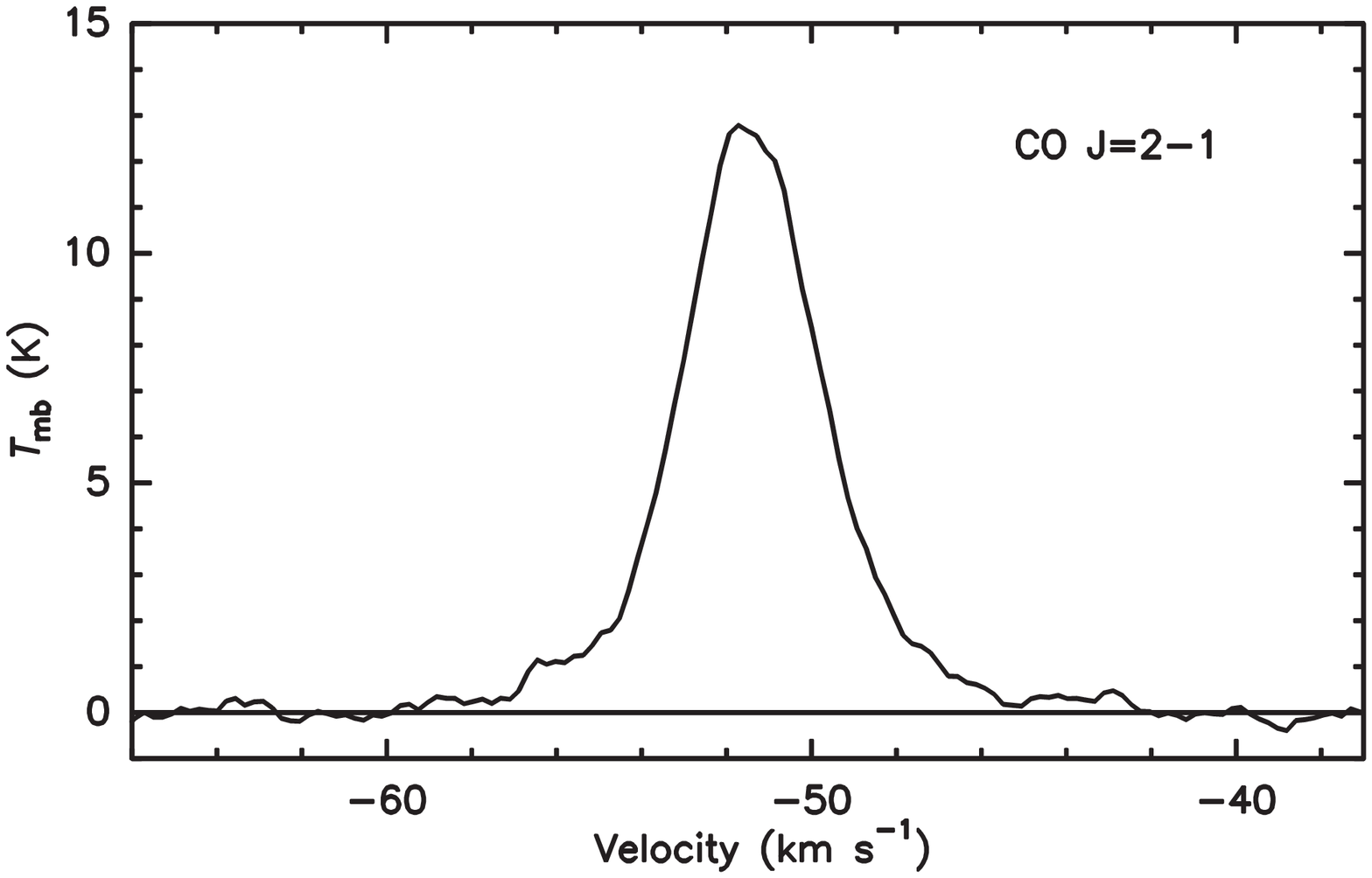}~~~~~%
   %{f1.co2-1.ps}%{fig1_1.ps}
   \includegraphics[width=55mm]{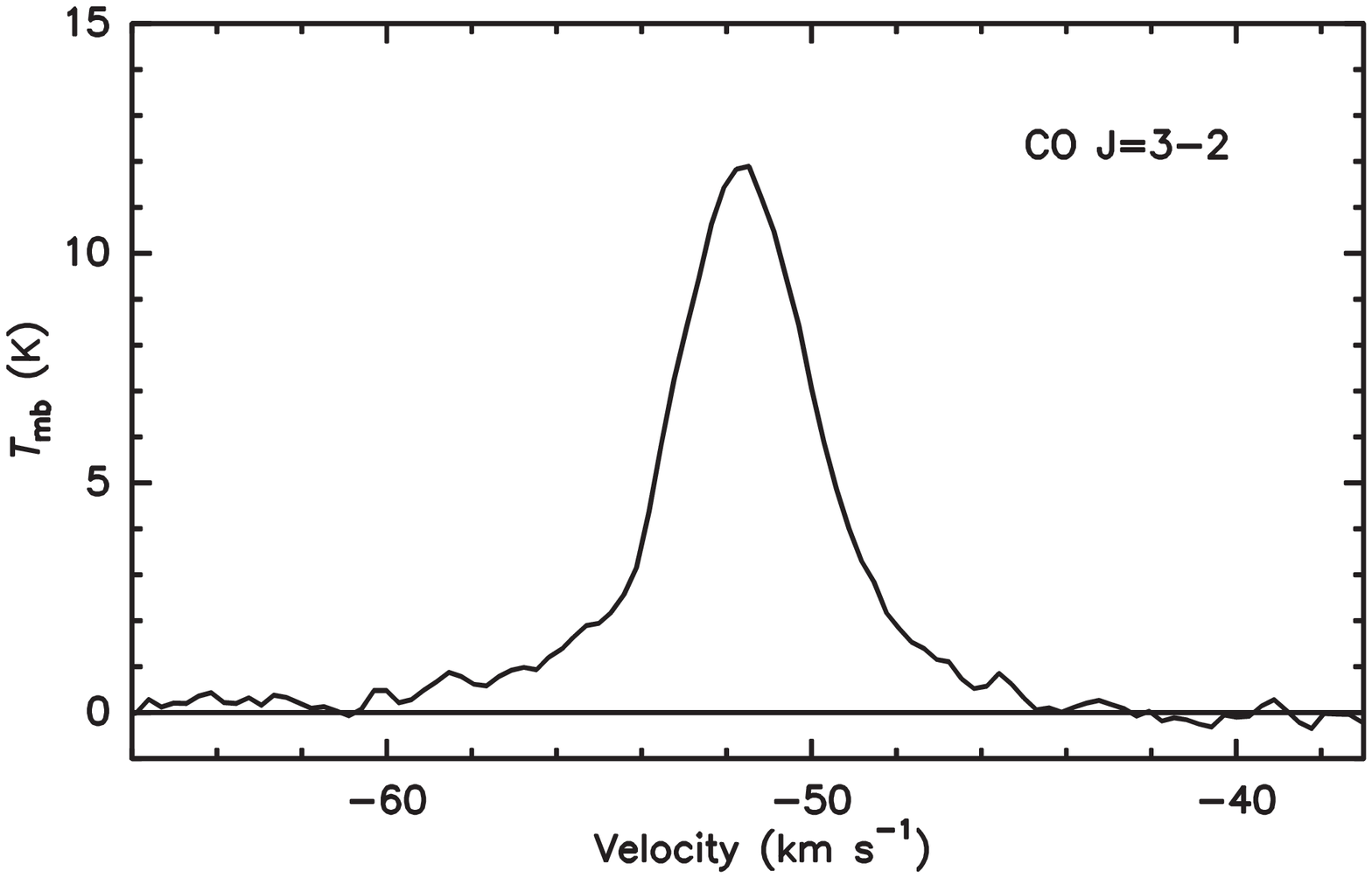}%{f1b.eps.co3-2.ps}%{fig1_2.ps}

\vs
  \includegraphics[width=55mm]{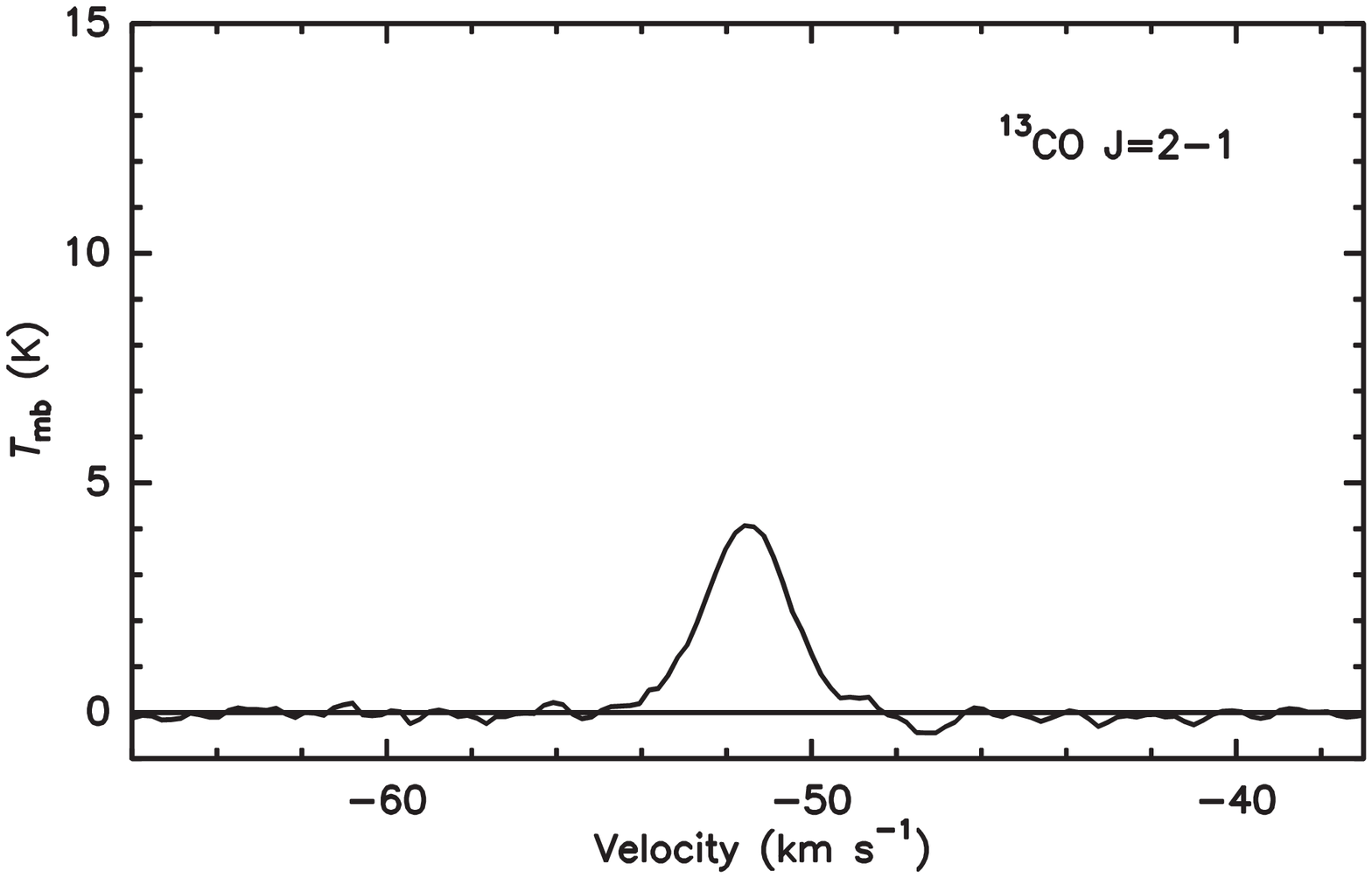}%{f1.13co2-1.ps}%{fig1_3.ps}
  \includegraphics[width=55mm]{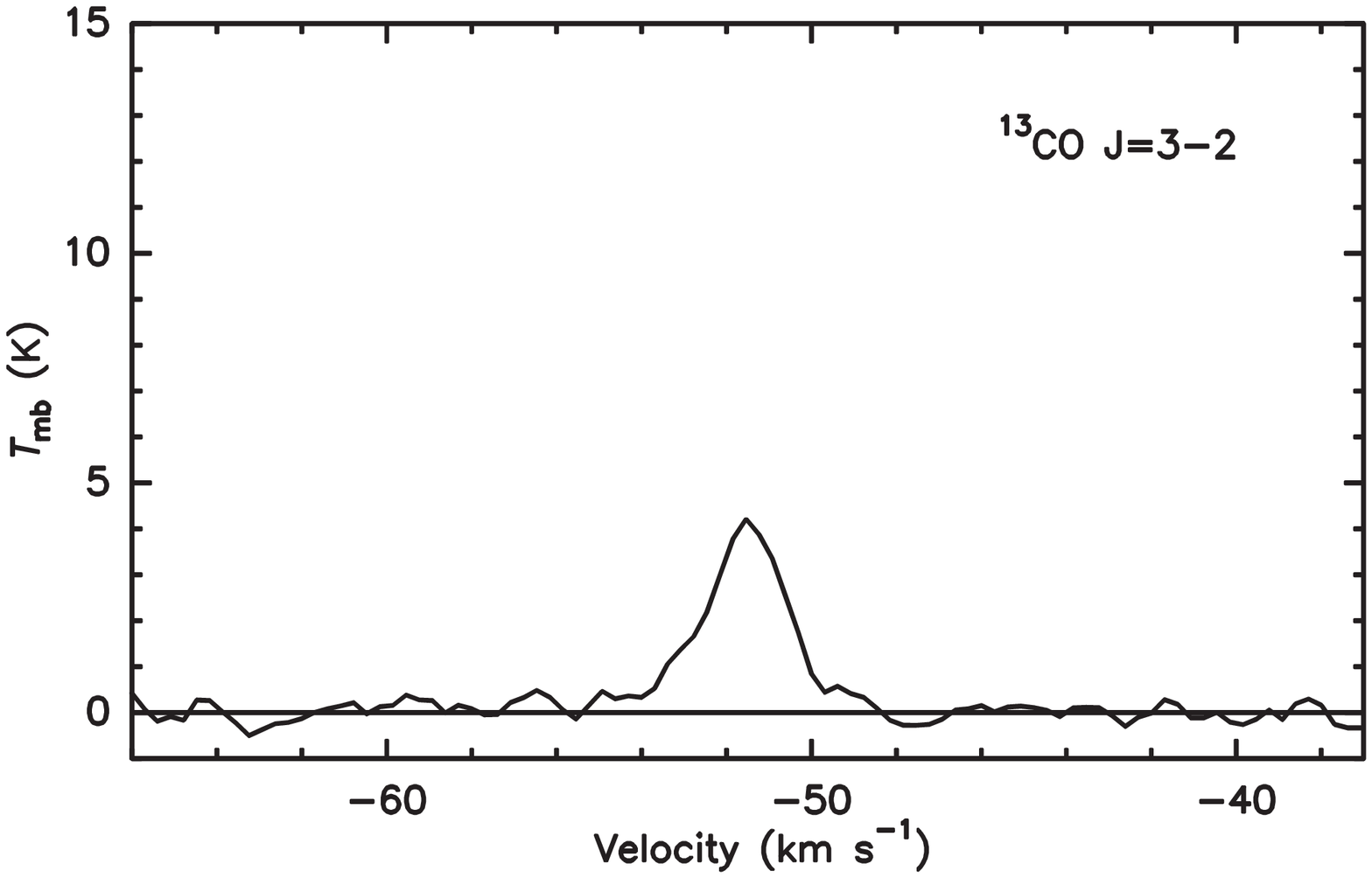}%{f1.13co3-2.ps}%{fig1_4.ps}

\vs

  \begin{minipage}{120mm}
  \caption{\baselineskip 3.6mm Spectra at the central position of the region (0, 0) of the mapping
  observations.}\end{minipage}
\end{figure}

\begin{table}[!h]
\centering \begin{minipage}{95mm} \caption[]{ Observational
Parameters of {the }IRAS~22506+5944 Source }\end{minipage}

\fns
  \begin{tabular}{lcccc}
  \hline\noalign{\smallskip}
Name   &$T_{\rm mb}$ & FWHM  & FW & $V_{\rm LSR}$                  \\
       &${\rm (K)}$&${\rm (km\ s^{-1})}$&${\rm (km\ s^{-1})}$&${\rm (km\ s^{-1})}$\\
\noalign{\smallskip}\hline\noalign{\smallskip}
CO $J=2-1$        & 12.8$\pm$0.1   & 3.98$\pm$0.03   & 11.11 & 51.43$\pm$0.01 \\  % new variable
CO $J=3-2$        & 11.9$\pm$0.1 & 4.15$\pm$0.06   & 13.63 & 51.59$\pm$0.02  \\  % new variable
$^{13}$CO $J=2-1$        &  4.1$\pm$0.1   & 2.37$\pm$0.04   & 6.69  & 51.52$\pm$0.02   \\  % new variable
$^{13}$CO $J=3-2$        &  4.0$\pm$0.3     & 2.32$\pm$0.08   & 7.22  & 51.54$\pm$0.01    \\
\noalign{\smallskip}\hline
\end{tabular}
\end{table}

\subsection{Dense Molecular Core}

In Figure~2, {all }the structures of {the} molecular cloud core
are presented and IRAS~22506+5944 has an isolated core. The cores
in {the} $J=2-1$ transition lines have a similar morphology to
{those} of {the} $J=3-2$ transition lines. Based on the results
observed by other authors, there {is} a near-infrared source S4
which was detected by Wang (1997) {as well as} three H$_{2}$O
masers in this region. The IRAS sources and the H$_{2}$O masers
are located closer to the core peak position traced by the
transition $J=2-1$, but a little deviation from that {is also}
traced by the transition $J=3-2$. The IRAS sources are associated
with the H$_{2}$O masers. %which
This indicates that {they are} still undergoing {activity in} an
early evolution stage (Felli et al. 1997). {The }H$_{2}$O masers
occur in relatively warm environments. Using the IRAS point-source
catalog, infrared luminosity (Casoli et al. 1986) and dust
temperature (Henning et al. 1990) are given by, respectively:
\begin{equation}
\mathit{L}_{\rm IR}=(20.653\times F_{12}+7.358\times
F_{25}+4.578\times F_{60}+1.762\times F_{100})\times
D^{2}\times0.30,
\end{equation}
\begin{equation}
\mathit{T}_{D}=\frac{96}{(3+\beta)\ln(100/60)-\ln(F_{60}/F_{100})},
\end{equation}
where $D$ is the distance from the sun in kpc{;} $F_{12}$,
$F_{25}$, $F_{60}$,{ and} $F_{100}$ are the infrared fluxes at
four IRAS bands ($12\,\mu$m, 25\,$\mu$m, 60\,$\mu$m and
100\,$\mu$m), respectively. The emissivity index of {the }dust
particle{s} ($\beta$) is assumed to be 2. The calculated results
are presented in Table~2. %222222222222222222

The core parameters are calculated following the procedure of Lee
et al. (1990). Assuming LTE, {we write }the column density as
\begin{equation}
\mathit{N}=2.4\times10^{14}\frac{\exp[hBJ(J+1)/kT_{\rm
ex}]}{J+1}\times\frac{(T_{\rm ex}+hB/3\,{\rm
K})\times\tau\times\Delta V}{1-\exp(-h\nu/kT_{\rm ex})} {\rm
cm}^{-2},
\end{equation}
where $B$ is the rotational constant of the molecule,  $J$ is the
rotational quantum number and $\nu$ is the frequency of {the}
spectral line. $\Delta V$ is a half full-width of speed, $T_{\rm
ex}$ is the excitation temperature and $\tau$ is the optical
depth. Here we assume $N$(H$_{2}$)/$N$(CO)=$10^{4}$ (Dickman
1978).

\begin{figure}[h!!]

\vs\vs \centering
  \includegraphics[width=55mm]{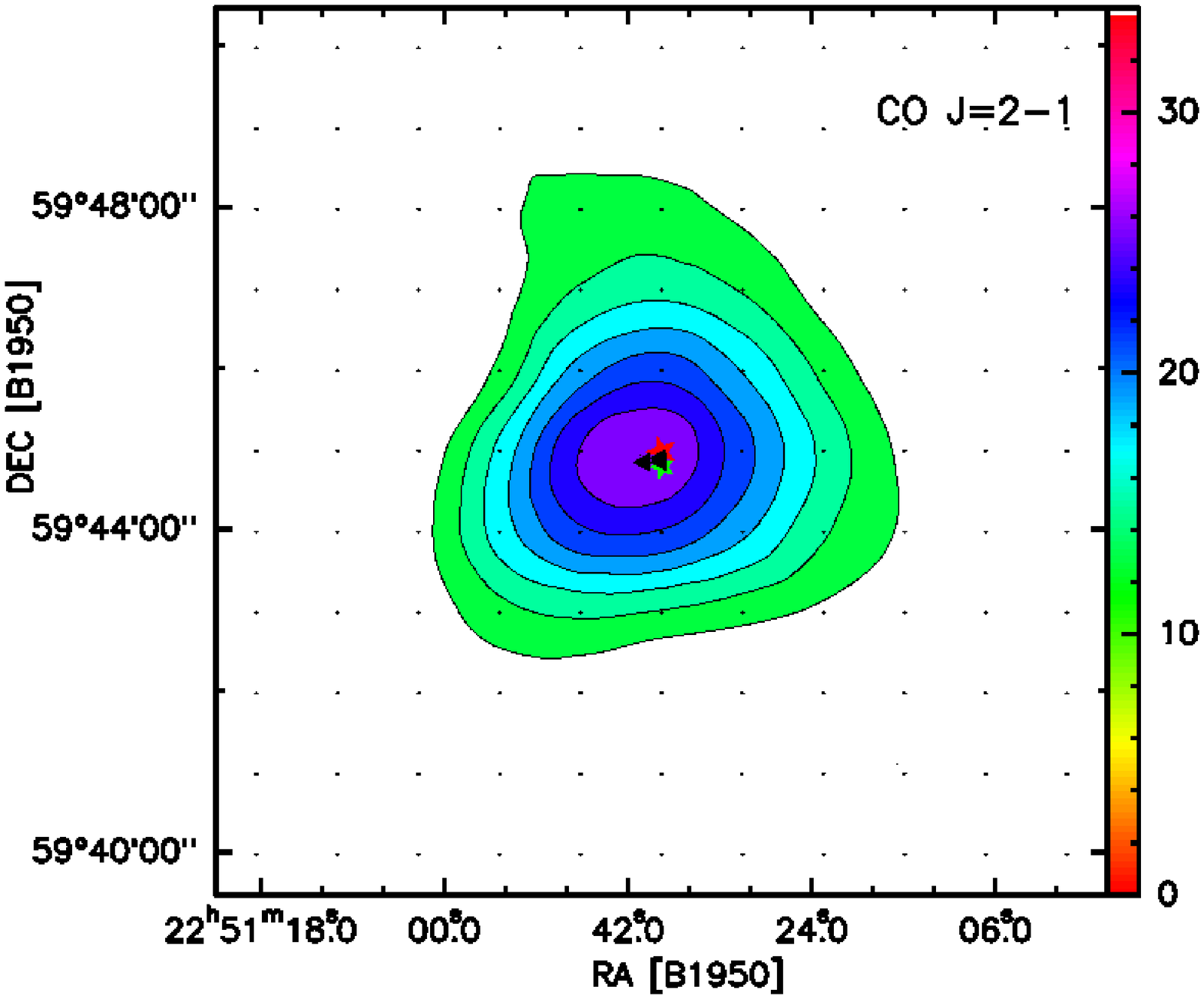}~~~~~~~%{f2.co2-1.ps}%{f2new.eps}~~~~~%{f2.co2-1.ps}
  \includegraphics[width=55mm]{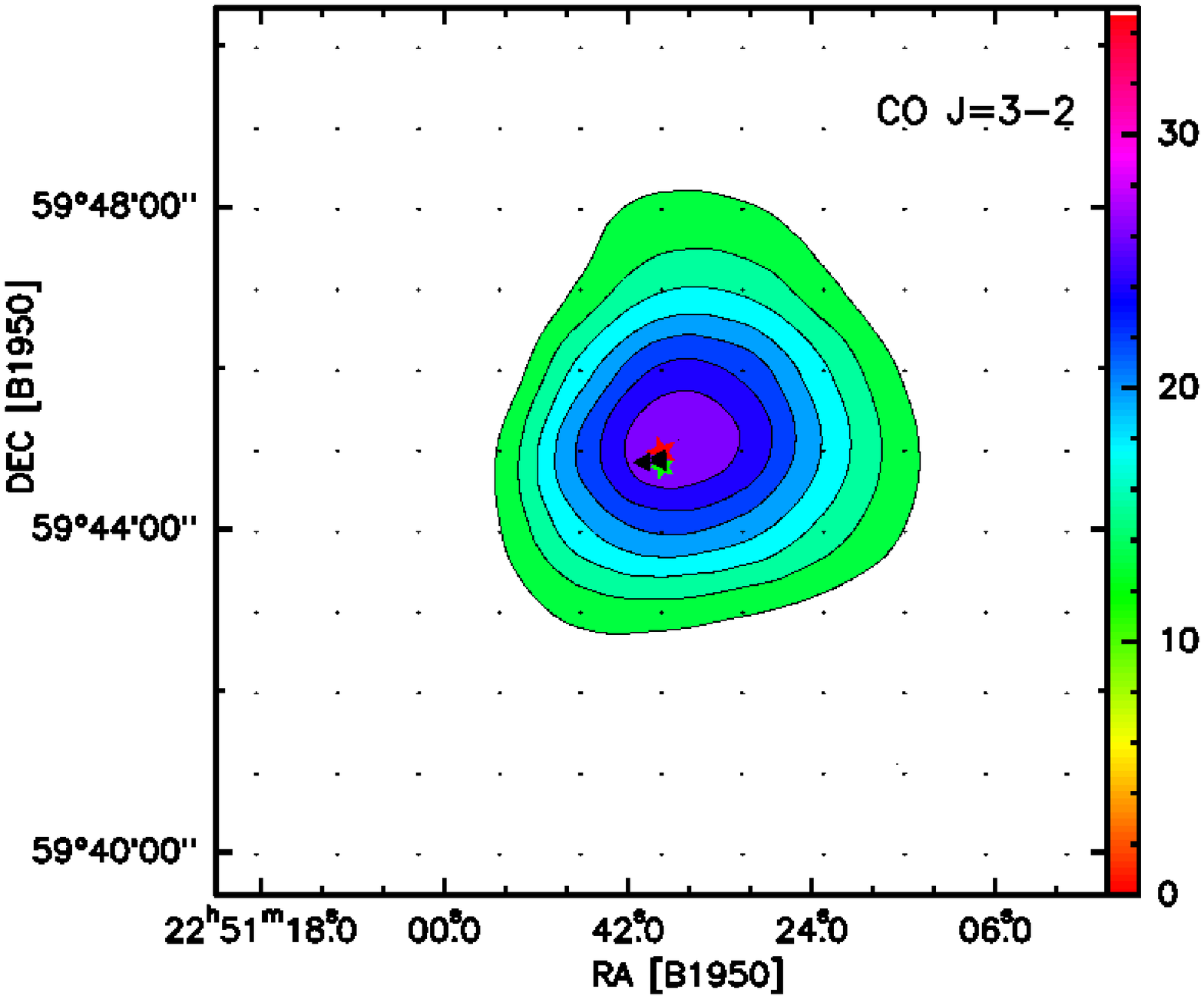}%{f2.co3-2.ps}

\vs\vs
  \includegraphics[width=55mm]{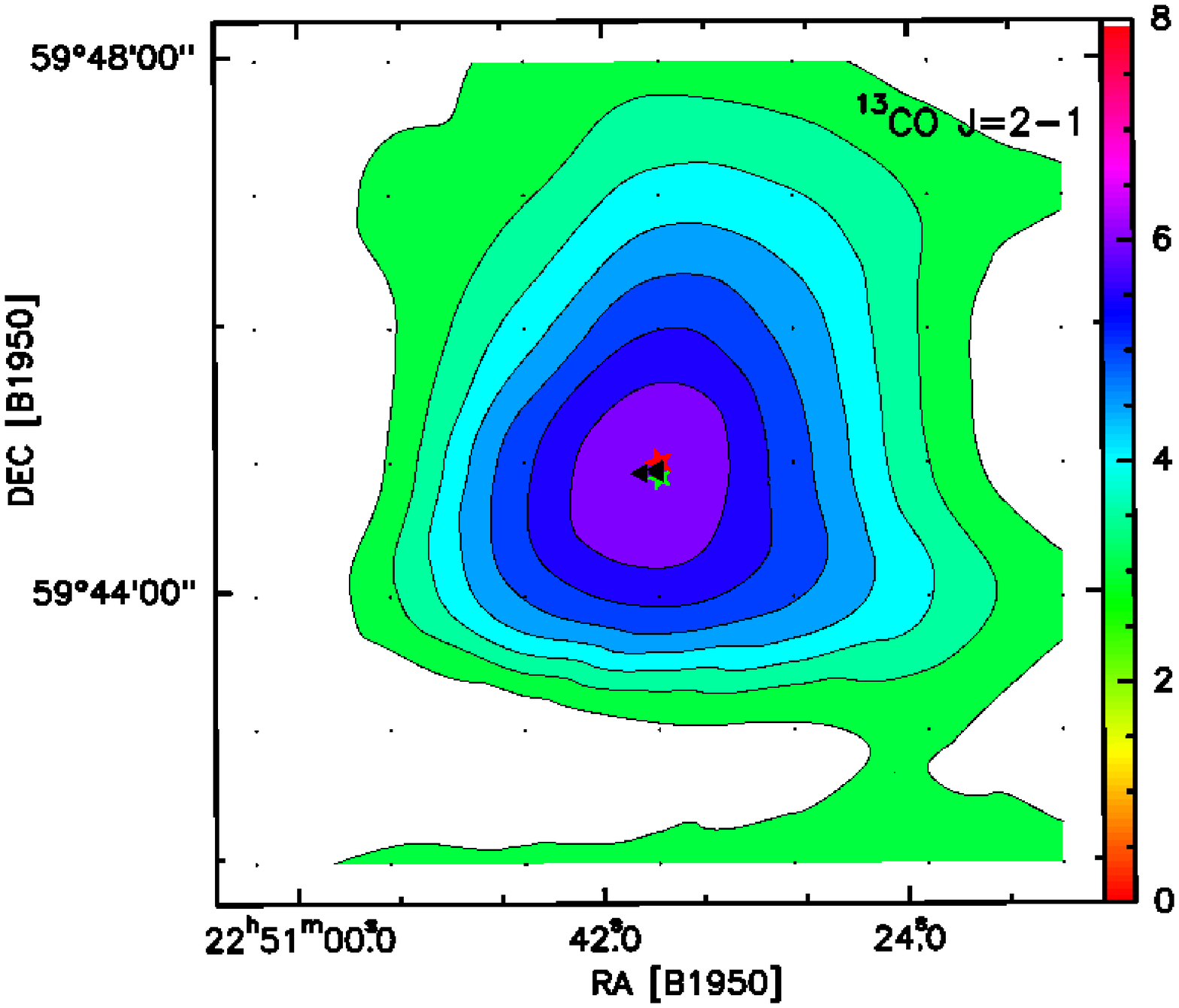}~~~~~~~%{f2.13co2-1.ps}
  \includegraphics[width=55mm]{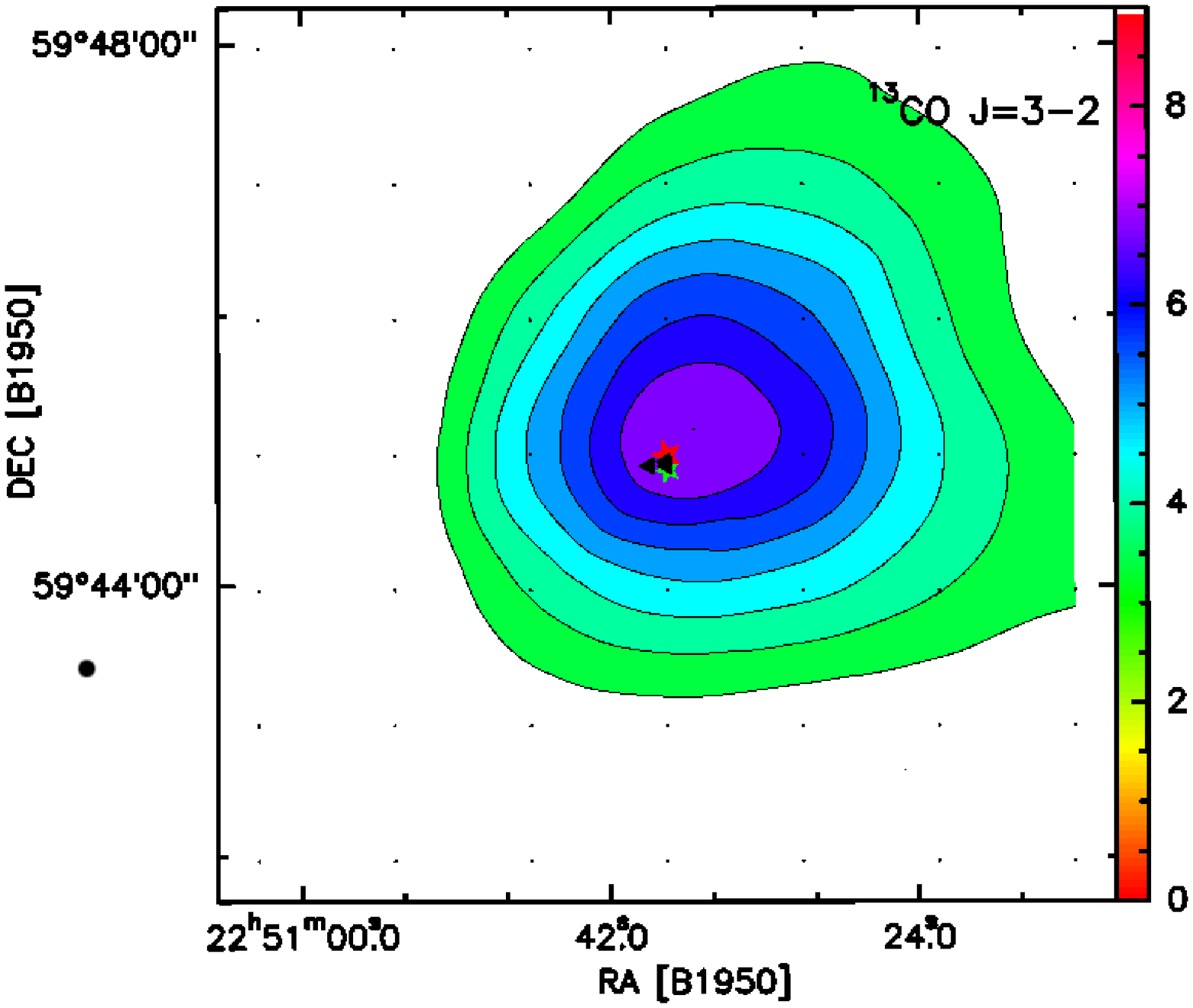}%{f2.13co3-2.ps}

\vs \caption{\baselineskip 3.6mm  Integrated intensity maps of the
core emission. In each map, the integrated range is from --53.26
to --49.68 ${\rm km\ s^{-1}}$. The contour levels are 30\%,
40\%,..., 90\% of the peak value. The red star represents {the}
IRAS~22506+5944 source position, the green star represents {the}
near-infrared source S4,{ and} the {filled} triangles indicate the
positions of the three H$_{2}$O masers. The dot symbols mark the
mapping points.}
\end{figure}

\begin{table}[h!!]
\centering \begin{minipage}{65mm}\caption[]{IRAS Flux{e}s and
Derived Parameters }
  \label{Tab:publ-works1}\end{minipage}

\vspace{-3mm}

  \fns\tabcolsep 1mm
 \begin{tabular}{lccccccccc}
  \hline\noalign{\smallskip}
Source& Distance &
$F_{12}$&$F_{25}$&$F_{60}$&$F_{100}$&$\lg(F_{25}/F_{12})$
&$\lg(F_{60}/F_{12})$&$T_{D}$&$L_{\rm IR}$      \\
&(kpc)&($J_{\nu}$)&($J_{\nu}$)&($J_{\nu}$)&($J_{\nu}$)&
& &(K)&($\times10^{4}L_{\odot}$)  \\
  \noalign{\smallskip}\hline\noalign{\smallskip}
IRAS 22506+5944  & 5.1&6.37&34.74&187.50&295.10&0.74&1.47&31.90 &
1.38 \\ \noalign{\smallskip}\hline
\end{tabular}
\end{table}

The CO emission is always considered optically thick while the
$^{13}$CO emission is usually optically thin. Therefore{,} the
excitation temperature $T_{\rm ex}$(CO) is given by (Garden et al.
1991):
\begin{equation}
\mathit{T}_{\rm ex}({\rm
CO})=\frac{hv}{k}\left({\ln\left\{1+\frac{hv}{k}{\left[\frac{T_{\rm
mb}}{f}+\frac{hv}{k}\left(\exp\Big(\frac{hv}{kT_{\rm
bg}}\Big)-1\right)\right]^{-1}}\right\}}\right)^{-1},
\end{equation}
where $T_{\rm bg}$=2.732\,K is the temperature of the cosmic
background radiation,{ and} $f$ is the beam filling factor. {In
addition, }$\tau(^{13}$CO) is given by (Garden et al.1991):
\begin{equation}
\ \tau(^{13}{\rm CO})=-\ln {\left\{1-\frac{kT_{\rm
mb}}{hv}\left[\frac{1}{\exp(hv/kT_{\rm
ex})-1}-\frac{1}{\exp(hv/kT_{\rm bg})-1}\right]^{-1}\right\}}.
\end{equation}
We also use another method to calculate $\tau(^{13}$CO):
\begin{equation}
\frac{T_{\rm mb}({\rm CO})}{T_{\rm mb}(^{13}{\rm
CO})}\approx\frac{1-\exp[-\tau({\rm CO})]}{1-\exp[-\tau(^{13}{\rm
CO})]}.
\end{equation}
We assume the solar abundance ratio
[CO]/[$^{13}$CO]=$\tau($CO)/$\tau(^{13}$CO)=89 (Lang 1980). The
calculated results are listed in Table~3. %333333333333333333
 From
the table, we can see that the optical depth{s} calculated by the
above two methods are almost equal. Thus, we suggest that the
abundance ratio [CO]/[$^{13}$CO] in this region is nearly the same
as{ that in} our solar system. Using the column density, the
mass{es} of outflow gas are obtained by:
\begin{equation}
\mathit{M}=\mu N_{\rm H_{2}}S/(2.0\times10^{33}),
\end{equation}
where the mean atomic weight of the gas $\mu$ is 1.36,{ and} $S$
is the size of {the} core region. The physical parameters of the
core are summarized in Table~4. %44444444444

\begin{table}[h!!]
\centering \begin{minipage}{70mm}
 \caption[]{ Calculated Results of the Optical Depth}
  \label{Tab:publ-works2}\end{minipage}

\vspace{-3mm} \fns\tabcolsep 5mm

\begin{tabular}{lcc}
  \hline\noalign{\smallskip}
Name   & $\tau_{1}$ &   $\tau_{2}$                \\
\noalign{\smallskip}  \hline\noalign{\smallskip}
CO  $J=2-1$        & 33.55 & 34.00\\  % new variable
CO  $J=3-2$        & 52.33 & 53.59  \\  % new variable
$^{13}$CO $J=2-1$ & 0.38 & 0.38 \\  % new variable
$^{13}$CO $J=3-2$ & 0.59 &  0.60 \\
\noalign{\smallskip}\hline\noalign{\smallskip}
\end{tabular}

\parbox{60mm}{The optical depth ($\tau_{1}$) is
  derived based on the first method. ($\tau_{2}$) is derived based on the second
  method.}
%\end{table}
%\begin{table}[]

\vs\vs \small

\centering \begin{minipage}{60mm} \caption[]{ Physical Parameters
of the Core }
  \label{Tab:publ-works3}\end{minipage}

\vspace{-3mm}
  \fns\tabcolsep 2mm
  \begin{tabular}{lccccc}
  \hline\noalign{\smallskip}
Name   & $T_{\rm ex}$ &$\tau$& $N$(CO $J=2-1$) & $N$(H$_{2}$)& $M$               \\
       &  (K)   &   & ($\times10^{17}$cm$^{-2}$) & ($\times10^{21}$cm$^{-2}$)&  ($\times10^{3}\,M_{\odot}$)                 \\
\noalign{\smallskip} \hline\noalign{\smallskip}
IRAS 22506+5944        & 20.5& 33.55 & 7.07 & 7.07 & 2.03\\  % new variable
\noalign{\smallskip}\hline
\end{tabular}
\end{table}

\subsection{The Bipolar Outflows}

From Figure~3, bipolar outflows {are} clearly revealed in {the}
IRAS~22506+5944 region. The red wing and blue wing lobes mostly
overlap, which may be attributed to the axis of outflows {being}
nearly parallel with the line of sight direction. The
position-velocity (PV) diagrams in Figure~4, with a cut along the
north-south direction, also clearly show high-velocity outflows.
The outflows of the CO $J=3-2$ are much clearer than {those} of
the CO $J=2-1$. Two IRAS sources are located close to the center
of the outflows, so they may be the outflows driving {the
}source{,} {b}ut we {cannot} distinguish which source is the
driving source.
%------------------------------------------------------------------------------- Figs 3 :
\begin{figure}%[h]
  \centering

  \vs
  \includegraphics[angle=-90,width=55mm]{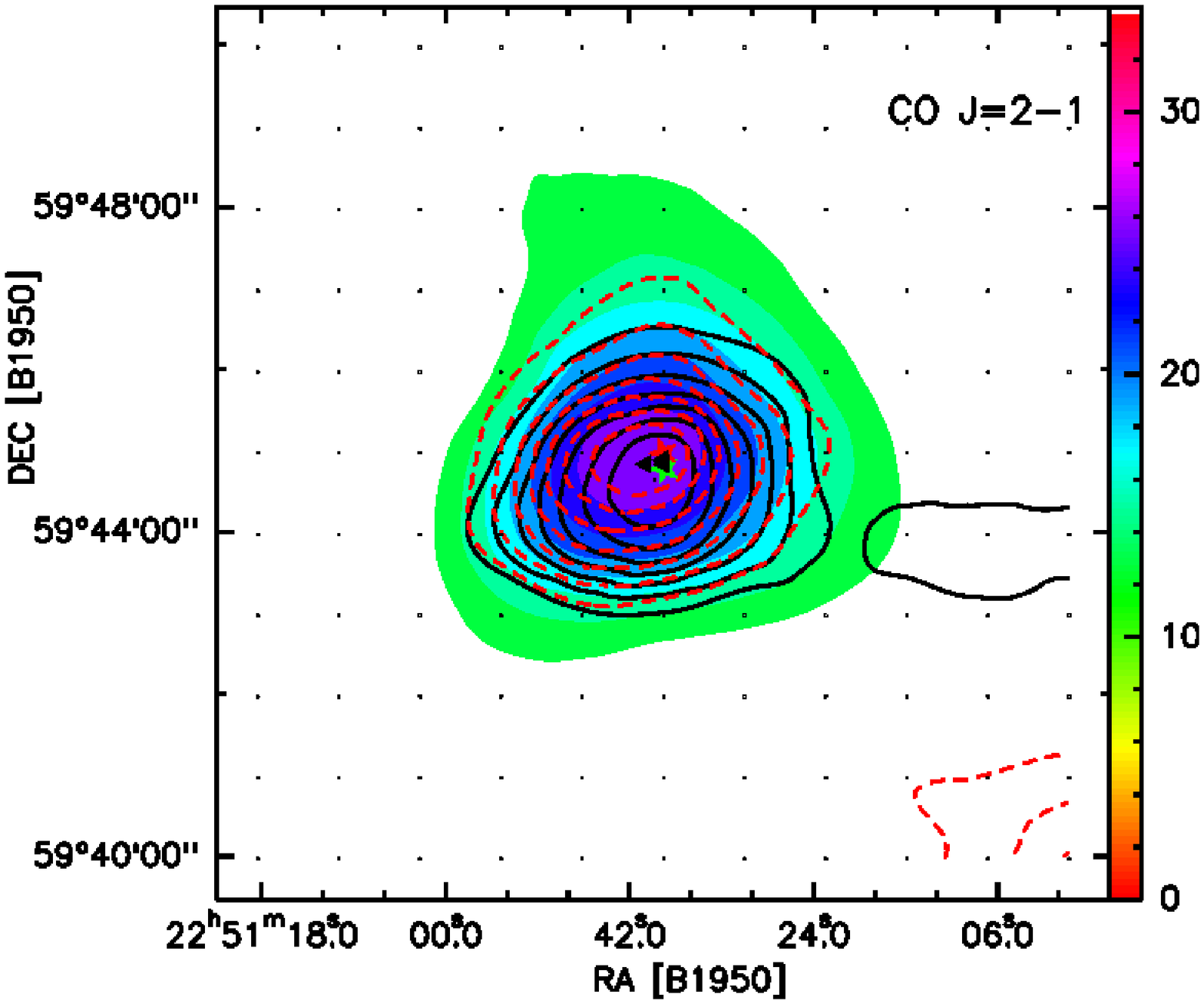}~~~~~
  \includegraphics[angle=-90,width=55mm]{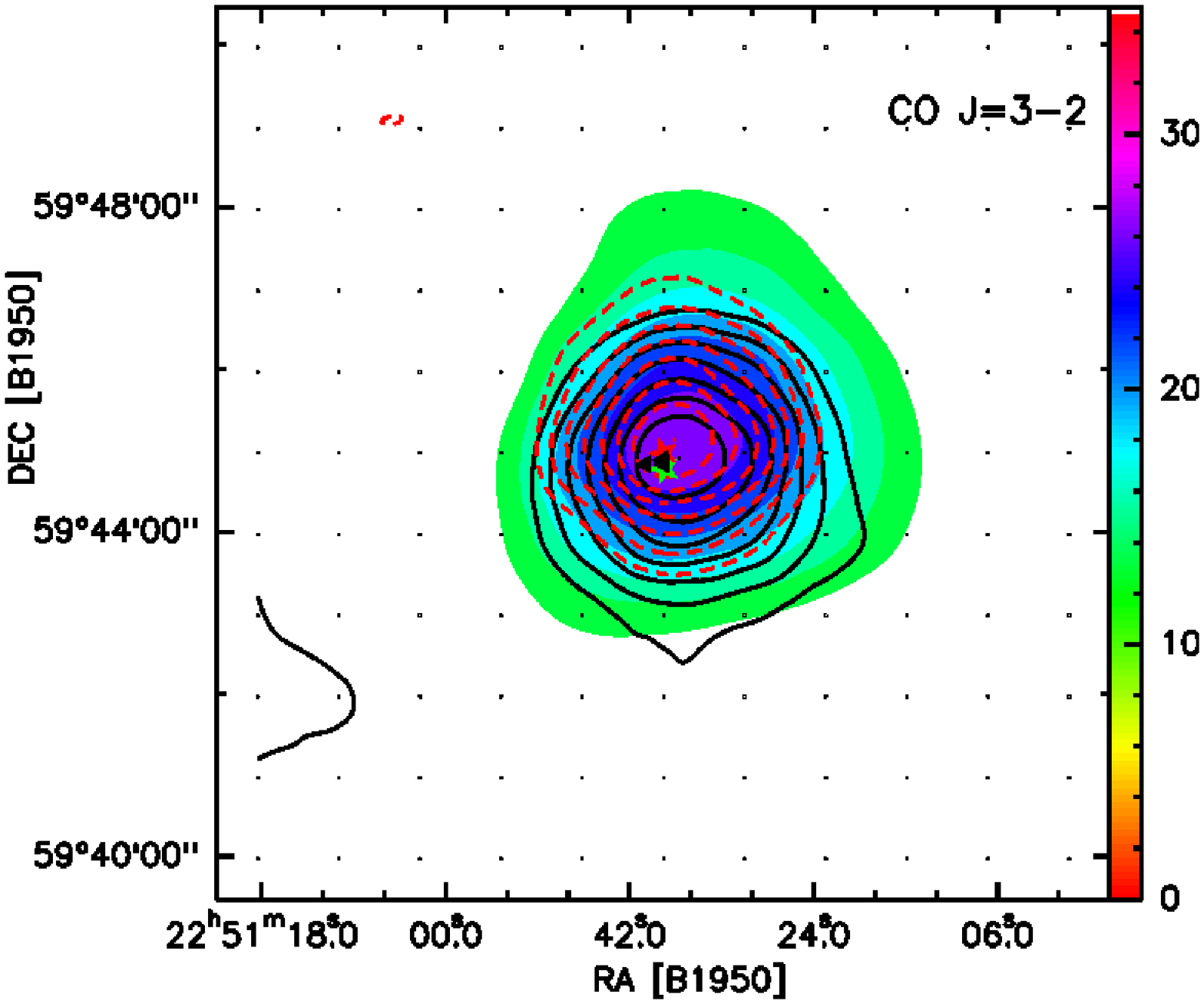}

  \caption{\baselineskip 3.6mm {\it Left panel}: outflow
contours of the CO $J=2-1$ are superimposed on the integrated
intensity map of the core{; the} integrated range is from --57.07 to
--53.26 ${\rm km\ s^{-1}}$ for the blue wing ({\it %blue
black solid line}) and from --49.68 to --45.96 ${\rm km\ s^{-1}}$
for the red wing ({\it red dashed line}). {\it Right panel}:
outflow contours of the CO $J=3-2$ are superimposed on the
integrated intensity map of the core{; the} integrated range is
from --58.96 to --53.26 ${\rm km\ s^{-1}}$ for the blue wing ({\it
%blue
black  solid line}) and from --49.68 to --45.33 ${\rm km\ s^{-1}}$
for the red wing ({\it red dashed line}). The contour levels for
core and outflows are 30\%, 40\%, ..., 90\% of the peak values. }
%\end{figure}
%----------------------------------------------------------------------------- Figs 4 :
%\begin{figure}%[h]
  \vspace{4mm} \centering
  \begin{minipage}[t]{0.500\linewidth}

  \includegraphics[angle=-90,width=55mm]{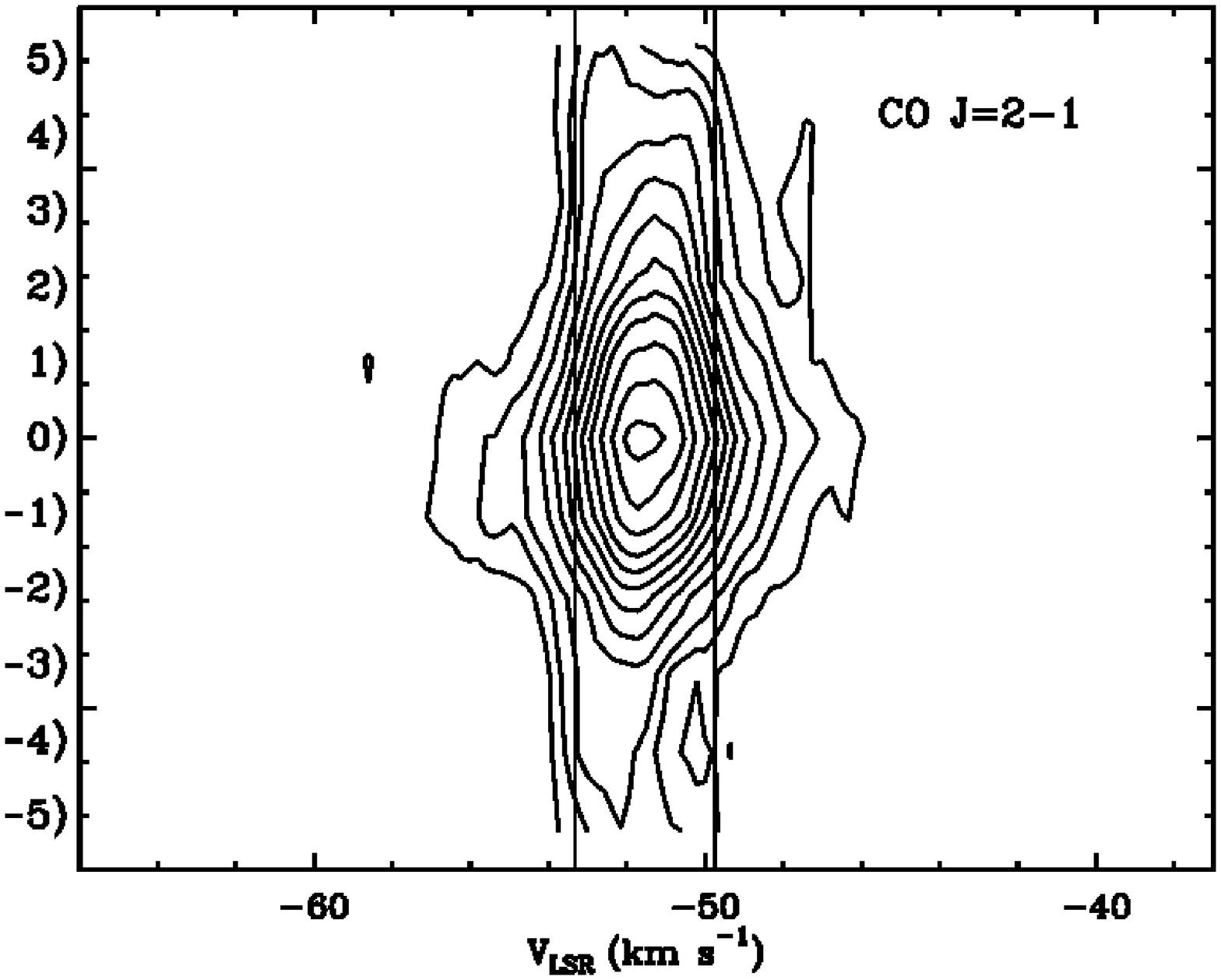}
  \vspace{-4mm}
  \end{minipage}%
  \begin{minipage}[t]{0.400\textwidth}
  \centering
  \includegraphics[angle=-90,width=55mm]{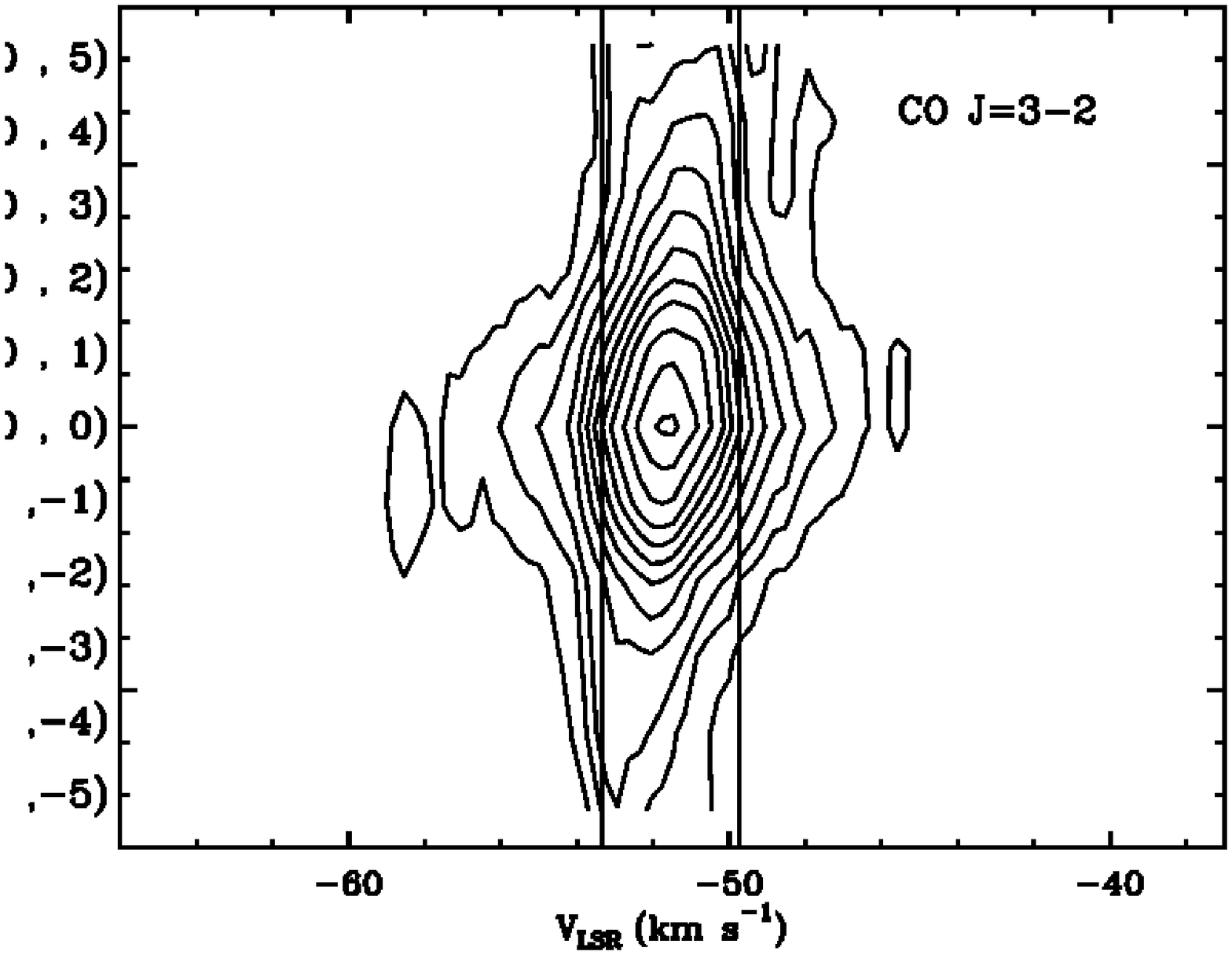}
  \vspace{-1mm}
  \end{minipage}%
  \begin{center}

  \caption{\baselineskip 3.6mm {\it Left panel}:
   P-V diagram constructed from the CO $J=2-1$ transition. Contour levels are 1, 2, 3,
   4.5, 6, 6.5, 7.5,
  9, 10.5, 12, 14,..., 25\,K.
  {\it Right panel}: P-V diagram for CO $J=3-2$. Contour levels are the same as in {the} left panel. The left and right vertical lines
indicate the beginning of the blue and red wings, respectively. }
  \end{center}
\end{figure}

Under {the} LTE assumption, the averaged column density of the
outflows can be obtained by (Scoville et al. 1986):
\begin{equation}
\mathit{N}=2.4\times10^{14}\frac{\exp[hBJ(J+1)/kT_{\rm
ex}]}{J+1}\times\frac{(T_{\rm ex}+hB/3\,{\rm K})}{\exp(-hv/kT_{\rm
ex})}\int T_{\rm mb}\times\frac{\tau dv}{[1-\exp(-\tau)]}{\rm
cm}^{-2}.
\end{equation}
We consider that the excitation temperature is uniform in the
observed region. The excitation temperature ($T_{\rm ex}$) is 20.5 K
from the calculated results of the cores. The $\tau$ can be
determined by Equation~(6). The integral is carried out in the
blue and red wing {regions} and the
integrated ranges are determined from the PV diagram.

In addition, we can derive the outflows{'} mass ($M$) from Equation~(7).
The momentum $P$ and energy $E$ are calculated by $P=MV$ and
$E=MV^{2}$, where $V$ is the mean velocity of the gas relative to
the cloud{'s} systemic velocity. A dynamic time scale can be determined
by $t_{\rm d}=R/V$, where $R$ is {the} mean size of the
outflows. The driving force is determined by $F=P/t_{\rm d}$. The
mechanical luminosity and the mass loss rate of the outflows are
calculated using $L_{\rm mech}=E/t_{\rm d}$ and
 $\dot{M}=P/(t_{\rm d}v_{\rm w})$, where the final wind velocity is taken to
be $v_{\rm w}=500\,{\rm km\ s^{-1}}$ (Marti et al. 1986). The
calculated results are listed in Table~5. % 5555555555555555

\begin{table}[h!!]

\vs \centering \begin{minipage}{65mm}\caption[]{ Physical
Parameters of the Outflows}
  \label{Tab:publ-works4}\end{minipage}

\vspace{-3mm}
  \fns\tabcolsep 0.3mm
  \begin{tabular}{lccccccccr}
  \hline\noalign{\smallskip}
Name & Wing & $N$(H$_{2}$)& $M$  & $t_{\rm d}$ &$\dot{M}$&$F$&$P$&$E$
&$L_{\rm Mech}$           \\
IRAS       & & ($\times10^{20}$\,cm$^{-2}$)&  ($M_{\odot}$)&
($\times10^{5}$\,yr)
       &($M_{\odot}$\,yr$^{-1}$)
       &($M_{\odot}$\,km~s$^{-1}$~yr$^{-1}$)&($M_{\odot}$\,km~s$^{-1}$)
        &($\times10^{46}$\,erg)&($L_{\odot}$)              \\
 \noalign{\smallskip} \hline\noalign{\smallskip}
 22506+5944   & Blue  & 1.35 & 18.7 & 4.14& $8.0\times10^{-7}$& $4.0\times10^{-4}$  & 164.6 & 1.08 & 0.22 \\  % new variable
                    & Red   & 1.35 & 18.6 & 4.38 &$8.2\times10^{-7}$& $4.1\times10^{-4}$  & 179.6 & 1.30 & 0.24\\
\noalign{\smallskip}\hline
\end{tabular}
\end{table}

\section{Discussion}
\label{sect:discussion}

From Figure~1 and Table~1, the full width of the CO $J=3-2$ line is
larger than that of the CO $J=2-1$ line and the line wings in
CO $J=2-1$ {are} broader than {those}
in CO $J=1-0$ (Wu et al. 2005). These results
suggest that the outflows may be caused by stellar wind sweeping up
the surrounding materials in different intensity layer{s} and
the outflows traced by the higher $J$ level CO lines arise
from the warm layer closer to the central exciting star.

Although the color indexes of the source satisfy the criteria of
Wood $\&$ Churchwell (1989), {they are} not detected in the
centimeter or millimeter continuum emission (Wu et al. 2005).
Whether or not it is a {UC HII} region still needs to be
determined by further higher sensitivity continuum observations.
So far{,} the majority of authors have used constant values of
$\tau$ and $T_{\rm ex}$ to estimate total mass. {Because} our
observations were simultaneously made in CO and $^{13}$CO, we can
obtain relatively accurate values of $\tau$ and $T_{\rm ex}$, as
well as {the} total mass. From Table~4, the total mass of the core
is 2.03$\times10^{3}\,M_{\odot}${;} the value indicates that
IRAS~22506+5944 is a high-mass star formation region. Wang (1997)
found clusters of stellar objects and an infrared jet in this
region. Thus{,} we conclude that IRAS~22506+5944 is associated
with molecular cloud complexes. The IRAS~22506+5944 source appears
to be {a} deeply embedded protostar.

In addition, the contour maps and PV diagram of the
IRAS~22506+5944 source {clearly }show bipolar outflows in this
region. The total mass of the outflows is 37.3\,$M_{\odot}$, {and
}the total{ amount of} momenta is 344.2\,$M_{\odot}$\,km~s$^{-1}$.
Both parameters are much larger than the typical values in
low-mass outflows. The larger outflow{'s} mass suggests that the
outflows{'} mass is not likely to originate from the stellar
surface, but could be caused by the entrainment of ambient gas
(Shepherd $\&$ Churchwell 1996). The IRAS~22506+5944 source is
considered as the outflow driving source by Su
et al. (2004), while Wu et al. (2005) consider %the
  S4
to be the outflow driving source. Two IRAS source{s} are located
{at} the geometric position of the driving source of the outflow,
{and} we also cannot identify which one is the driving source. We
need to apply for observations {with a} much higher ang{u}l{ar}
resolution telescope to confirm it. $L_{\rm IR}/L_{\rm
Mech}\gg1$,{ and} $L_{\rm IR}/CF<1$, where $L_{\rm Mech}$ is the
total mechanical luminosity, suggest{ing} that despite the strong
radiation from the cent{ral} source, the radiation pressure still
cannot supply enough force to drive such massive outflows (Wu et
al. 1998).

\section{Conclusions}
\label{sect:conclusion} We  observed the CO $J=2-1$, CO $J=3-2$,
$^{13}$CO $J=2-1$ and $^{13}$CO $J=3-2$ lines {in the direction
of} IRAS~22506+5944. IRAS~22506+5944 has an isolated core. The
infrared data indicate that the IRAS~22506 +5944 source appears to
be {a} deeply embedded protostar. Bipolar outflows are {further
}identified in th{is} region. IRAS~22506+5944 or NIR source S4 may
be the outflow driving source. The two IRAS sources are associated
with H$_{2}$O masers. {The }H$_{2}$O masers occur in relatively
warm environments. The parameters of the core and the outflows are
derived{;} the derived values suggest that IRAS~22506 +5944 is a
high-mass star formation region and the {total }mass of the
outflows is larger than that in low-mass star formation region{s}.
Compared with the extension of line wings in CO $J=2-1$, CO
$J=3-2$ line wings extend further in velocity, suggesting that the
outflows traced by the higher $J$ level $^{12}$CO lines arise from
the warm layer closer to the central exciting star.

\normalem
\begin{acknowledgements}
We would like to thank Dr{s}. Sheng-Li Qin and Martin Miller for
{their} help {with} data
acqui{sition} and discussion. We are also grateful
to the referee for his/her helpful comments. This work was supported
by the National Natural Science Foundation of China (Grant
No.~10473014).
\end{acknowledgements}


\begin{thebibliography}{99}
\small \setlength{\itemindent}{-3mm} \setlength{\itemsep}{-0.5mm}
\setlength{\baselineskip}{4.7mm}


  \bibitem[1986]{Casoli86} Casoli, F., Combes, F., Dupraz, C., Gerin, M., \& Boulanger, F.  1986, A\&A, 169,
  281
  \bibitem[1978]{Dickman78} Dickman, R. L.  1978, \apj, 37, 407
  \bibitem[1989]{Elitzur89} Elitzur, M., Hollenbach, D. J.,  \& McKee, C. F.  1989, \apj, 346,
  983
  \bibitem[1997]{Felli97} Felli, M., Testi, L., Valdettaro, R., \& Wang, J. J.  1997, A\&A, 320,
  594
  \bibitem[1991]{Garden91} Garden, P. R., Hayashi, M., Hasegawa, T., Gatley, I., \& Kaifu, N.  1991, \apj, 374,
  540
  \bibitem[1990]{Henning90} Henning, Th., Pfau, W., \& Altenhoff, W. J.  1990,
  A\&A, 227, 542

  \bibitem[1995]{Jenness95} Jenness, T., Scott, P. F., \& Padman, R.   1995, MNRAS, 276,
  1024

  \bibitem[1993]{Lada93} Lada, E. A., Strom, K. M., \& Myers, P. C.  1993,
  Protostars and Planets III, eds. E. H. Levy, \& J. I. Lunine (Tucson:
  Univ. Arizonna Press), 245

  \bibitem[1980]{Lang80} Lang, K. R.  1980 A{s}trophysical {F}ormulae
   (Berlin: Springer-Verlag)

\bibitem[1990]{Lee90} Lee, Y., Snell, R. L., \& Dickman, R. L.  1990 \apj, 355, 536

  \bibitem[1990]{Marti90} Marti, J., Rodriguez, L. F.,
   \& Reipurth, B.  1998 \apj, 502, 377

  \bibitem[1991]{Molinari91} Molinari, S., Brand, J., Cesaroni, R., \& Palla, F.  1998 A\&A, 308,
  573

  \bibitem[1986]{Scoville86} Scoville, N. Z., Sargent, A. I., Sanders, D. B., et al.   1986, \apj, 303, 416


  \bibitem[1989]{Shepherd89} Shepherd, D. S., \& Churchwell, E.  1989,  \apj, 340,
  265


  \bibitem[1996]{Shepherd96} Shepherd, D. S., \& Churchwell, E.  1996,  \apj, 472,
  225
  \bibitem[1987]{Shu87} Shu, F. H., Adams, F. C., \& Lizano, S.   1987, ARA\&A,
  25, 23


  \bibitem[1998]{Su04} Su, Y.-N., Zhang, Q., \& Lim, J.  2004,  \apj, 604, 258

  \bibitem[1997]{Wang97} Wang, J. J. 1997, Ph.D.{T}hesis, Beijing
  Astron. Obs.

  \bibitem[1999]{Wood99} Wood, D. O. S., \& Churchwell, E.  1989, \apj, 340,  265

  \bibitem[1998]{Wu98} Wu, Y., et al.  1998, A\&AS, 18, 243

  \bibitem[2005]{Wu05} Wu, Y., Zhang, Q., Chen, H., Yang, C., Wei, Y., et al.  2005, \aj, 129, 330

  \bibitem[2001]{zhang01} Zhang, Q., Hunter, T. R., Brand, J., et al.  2001, \apj, 552, 167

\end{thebibliography}
\end{document}